\documentclass[final]{elsarticle} % This was 12pt
\usepackage{amssymb}
\usepackage{amsthm}
\usepackage{amsmath}
\usepackage{color}
\usepackage{color}
\usepackage{graphicx}

\usepackage{caption}
\usepackage{subfig}
\usepackage{tikz}
\usetikzlibrary{arrows}

\usepackage{stmaryrd}
\journal{Journal of Computational Physics}

\begin{document}

\begin{frontmatter}


\title{Adjoint characteristic decomposition \\of one-dimensional waves}
%The connection between continuous and discrete adjoints in wave propagation
\author{Luca Magri$^{a}$}
\address[label1]{Cambridge University Engineering Department, Trumpington Street, CB2 1PZ, Cambridge, United Kingdom}
\begin{abstract}
Adjoint methods enable the accurate calculation of the sensitivities of a quantity of interest.  The sensitivity is obtained by solving the adjoint system, which can be derived by {\it continuous} or {\it discrete} adjoint strategies. 
In acoustic wave propagation, continuous and discrete adjoint methods have been developed to compute the eigenvalue sensitivity to design parameters and passive devices  (\textit{Aguilar, J. G. et al, 2017, J. Computational Physics, vol. 341, 163--181}). 
In this short communication, it is shown that the continuous and discrete adjoint characteristic decompositions, and Riemann invariants, are connected by a similarity transformation. The results are shown in the Laplace domain. 
The adjoint characteristic decomposition is applied to a one-dimensional acoustic resonator, which contains a monopole source of sound. The proposed framework provides the foundation to tackle larger acoustic networks with a discrete adjoint approach, opening up new possibilities for adjoint-based design of problems that can be solved by the method of characteristics. 
\end{abstract}
\end{frontmatter}
%%%%%%%%%%%%%%%%%%%%%%%%%%%%%%%
%%%%%%%%%%%%%%%%%%%%%%%%%%%%%%%
\section{Introduction}
In acoustic wave propagation, the primitive variables, e.g., the acoustic velocity and pressure, can be decomposed by the method of characteristics. The solutions of the characteristic equations are the Riemann invariants, which are the acoustic travelling waves. 
On the one hand, it was shown that the discrete adjoint approach is very straightforward to implement because it only requires the complex transpose of the matrix applied to the Riemann invariants~\cite{Aguilar2017}. 
The sensitivity of the eigenvalue to the design parameters, which is the quantity of interest, is computed to machine precision. 
However, with the discrete adjoint approach, it is still an open question how to calculate the adjoint primitive variables, which contain the spatial information, from the adjoint Riemann invariants. 
On the other hand, the continuous adjoint approach enables the calculation of the adjoint primitive variables from the adjoint Riemann invariants, i.e., no spatial information is lost. The downside of the continuous adjoints in wave propagation is that the sensitivities are not guaranteed to be accurate to machine precision, and more involved mathematical derivation is required even in simple systems~\cite[see Secs. 2.4 and 5.4 in][]{Aguilar2017}. 
As reviewed in~\cite{Magri2019_amr}, the spatial information contained in the adjoint primitive variables is crucial for the design of acoustic resonators, for example, by 
(i) optimal placement of passive devices, such as acoustic dampers, to suppress loud oscillations; 
(ii) optimal change of the shape of the acoustic resonator to prevent loud oscillations from occurring;  and 
(iii) computation of the sensitivity to the acoustic boundary conditions. 
In this paper, the mathematical connection that relates the characteristic decompositions, and Riemann invariants, of the continuous and discrete adjoint systems is found. Therefore, the spatial dependence of the solution is recovered in the discrete adjoint approach. 
\section{Wave propagation with a monopole source of sound}
\begin{figure}[!t]
  \begin{center}  
 \includegraphics[width=0.6\textwidth]{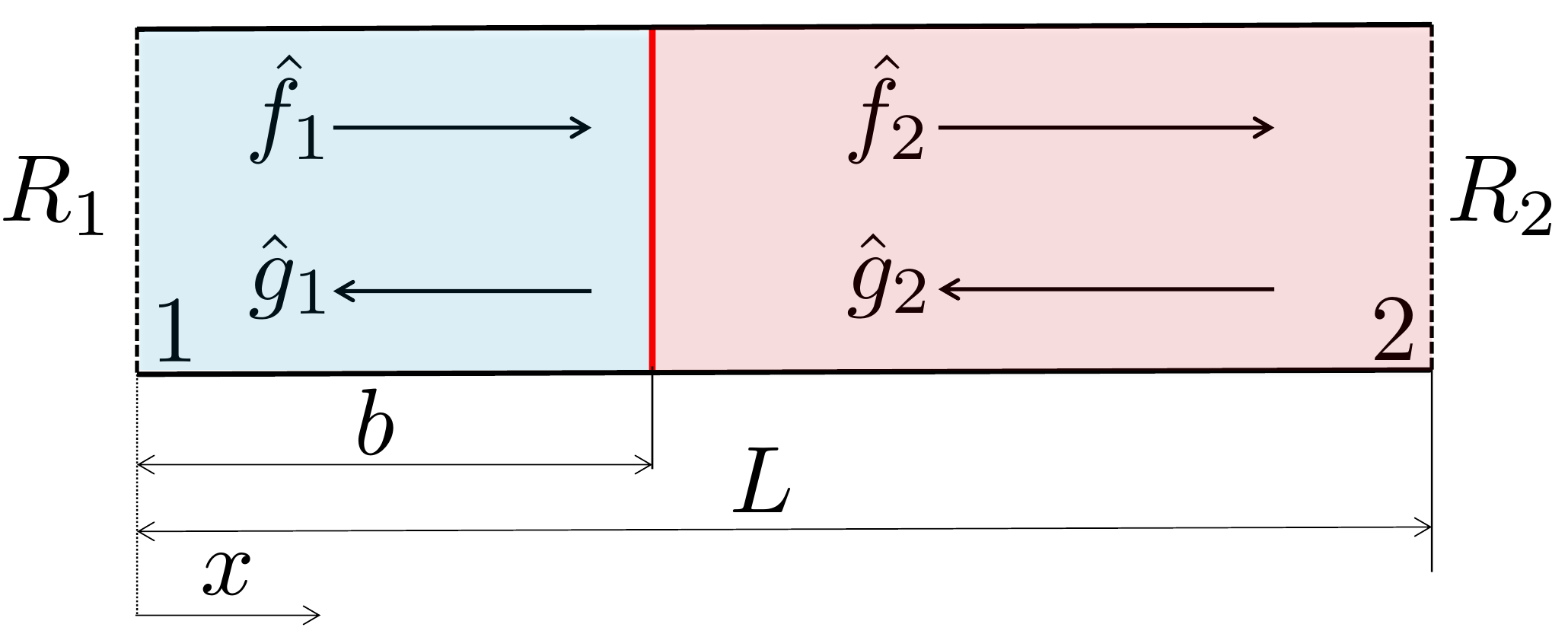}
  \end{center}
  \caption{Wave propagation and characteristic decomposition in an acoustic system with a compact heat source (vertical red solid line). Left (blue) / right (red) side is upstream / downstream of the flame.}\label{fig:ductSketch}
  \end{figure}

An acoustic resonator with a monopole source of sound, i.e., a flame,  is considered as a prototypical model of a thermoacoustic system (Fig.~\ref{fig:ductSketch}). 
The main assumptions of the model are~\cite{Aguilar2017}
(i) the ratio between the radius and length of the duct is sufficiently small such that only low-frequency longitudinal acoustics propagate (the axial coordinate is $x$);
(ii) the length of the flame is sufficiently smaller than the acoustic wavelength  such that
the heat is released at $x=b$ (compact assumption); 
(iii) the acoustics evolve on a slow mean flow such that the mean-flow Mach number is negligible; 
(iv) the flow is isentropic, except at the flame location; and 
(v)  the gas is ideal, $p=\rho\mathcal{R}T$, where $p$ is the pressure, $\rho$ is the density, $T$ is the temperature, and $\mathcal{R}$ is the gas constant. 
The speed of sound is $c=\sqrt{\gamma\mathcal{R}T}$, where $\gamma$ is the heat capacity ratio. For more details the reader may refer to~\cite{Aguilar2017}. 
The problem is governed by the nonlinear Euler equations with a model for the flame. 
Because the focus is on the acoustics of the resonator, which are considered as linear unsteady perturbations, a generic flow variable is decomposed as 
$(\cdot)=\bar{(\cdot)}+(\cdot)'$, 
where 
$\bar{(\cdot)}\sim\mathcal{O}(1)$ 
is the steady mean flow component, and 
$(\cdot)'\sim\mathcal{O}(\epsilon)$, 
with $\epsilon\rightarrow 0$, is the acoustic fluctuation.  
% Mean flow 
By substituting this decomposition in the nonlinear Euler equation, the mean-flow and acoustic equations are derived. 
On grouping the terms $\sim\mathcal{O}(1)$, 
the mean flow is fully characterized by algebraic equations that link the quantities upstream and downstream of the heat source.  
Two thermodynamic variables upstream of the heat source, e.g., $\bar{T}_1$ and $\bar{p}_1$, need to be set along with the increase in the mean-flow temperature, $\Delta\bar{T}$, induced  by the heat source, such that $\bar{T}_2=\bar{T}_1+\Delta\bar{T}$.   
The mean-flow pressure is constant in a zero-Mach number flow, i.e., $\bar{p}_1=\bar{p}_2\equiv\bar{p}$. 
The acoustic equations are obtained by grouping the variables $\sim\mathcal{O}(\epsilon)$. The linearized momentum and energy equations read, respectively  
\begin{subequations}
\begin{align}
\frac{\partial{u}'}{\partial t} + \frac{1}{\bar{\rho}}\frac{\partial  {p}'}{\partial  x} & =0,\label{eq:tacousticpdes1}\\
\frac{\partial {p}'}{\partial t} +\gamma \bar p \frac{\partial  {u}'}{\partial  x} & =0,  \label{eq:tacousticpdes2}
\end{align}
\end{subequations}
where $t$ is the time; $u'$ and $p'$ are the acoustic velocity and pressure, respectively. 
The equations are studied in the Laplace domain by modal decomposition 
$(\cdot)'(x,t)=\hat{(\cdot)}(x)\exp(st)$, 
where $s$ is a complex number, yielding 
\begin{subequations}
\begin{align}
s \hat{u} + \frac{1}{\bar{\rho}}\frac{d \hat{p}}{d x} & =0,\label{eq:acousticpdes1}\\
s\hat{p} +\gamma \bar p \frac{d \hat{u}}{d x} & =0.  \label{eq:acousticpdes2}
\end{align}
\end{subequations}
In compact notation, Eqs.~\eqref{eq:acousticpdes1}-\eqref{eq:acousticpdes2} are represented by 
\begin{align}
\mathcal{N}(s)\underline{\hat{q}} = 0, \label{eq:compactN}
\end{align}
where $\mathcal{N}$ is the linear differential operator that encapsulates the boundary conditions, and $\underline{\hat{q}}=[\hat{u},\hat{p}]^T$. 
Equations~\eqref{eq:acousticpdes1}-\eqref{eq:acousticpdes2} hold at any acoustic location except at the heat source, i.e., $x\in(0,L) - \{b\}$. 
At the heat-source location, $x=b$, the linearized momentum and energy equations in integral form provide the jump conditions~\cite{Aguilar2017}
\begin{align} 
\llbracket \hat{p} \rrbracket^{b^+}_{b^-} = 0, \label{eq:jump1}\quad\quad
\llbracket \hat{u} \rrbracket^{b^+}_{b^-} = \frac{\gamma-1}{\gamma\bar{p}} \hat{\mathcal{Q}}, 
\end{align}
where $\llbracket\cdot\rrbracket^{b^+}_{b^-}$ is the difference between the quantity $(\cdot)$ evaluated downstream of the flame, at $b^+$, and upstream of the flame, at $b^-$. 
The heat-release is $\hat{\mathcal{Q}}$. 
The flame is assumed to be perfectly premixed, such that the heat is released after a time delay, $\tau$, after that the acoustic velocity has perturbed the flame's base, i.e.,  $\hat{\mathcal{Q}} = n \hat{u}(b^-)\exp(-s\tau)$, where $n$ is the flame index. This is similar to the $n-\tau$ model proposed by Crocco~\cite{Crocco1969}, which captures the essential time-delayed physics of thermoacoustic instabilities. 
For $x\in(0,L)-\{b\}$, the primitive variables in Eqs.~\eqref{eq:acousticpdes1}-\eqref{eq:acousticpdes2} are decomposed in Riemann invariants through a characteristic decomposition as 
\begin{align}
\begin{bmatrix}
\hat{p}_i \\ \hat{u}_i
\end{bmatrix}
=
\begin{bmatrix}
\exp\left(-s\frac{x-b}{\bar c_i}\right)   &   \exp\left(s\frac{x-b}{\bar c_i}\right) \\
\frac{1}{\bar\rho_i \bar c_i}\exp\left(-s\frac{x-b}{\bar c_i}\right)   & \frac{-1}{\bar\rho_i \bar c_i}\exp\left(s\frac{x-b}{\bar c_i}\right)
\end{bmatrix}  
\begin{bmatrix}
\hat{f}_i \\ \hat{g}_i
\end{bmatrix}\label{eq:wave1}
\end{align}
where $i=1$ for a quantity upstream of the heat source, and 
$i=2$ for a quantity downstream of the heat source. 
The Riemann invariants $\hat{f}_i$ are the right-travelling waves, while the Riemann invariants $\hat{g}_i$ are the left-travelling waves (Fig.~\ref{fig:ductSketch}).
To connect the spatial dependency of the continuous and discrete adjoints (Sec.~\ref{sec:dacacon}), it is convenient to re-arrange Eq.~\eqref{eq:wave1} as
\begin{align}\label{eq:bigPhinonad}
\underbrace{
\begin{bmatrix}
\hat{p}_1 \\ \hat{u}_1 \\ \hat{p}_2 \\ \hat{u}_2
\end{bmatrix}
}_{\boldsymbol{\hat{\mathbf{q}}}(x)}
= 
\underbrace{
\begin{bmatrix}
\exp(s\frac{x-b}{\bar{c}_1})   &   0 & \exp(-s\frac{x-b}{\bar{c}_1}) & 0 \\
-\frac{1}{\bar{\rho}_1\bar{c}_1}\exp(s\frac{x-b}{\bar{c}_1})   &   0 & \frac{1}{\bar{\rho}_1\bar{c}_1}\exp(-s\frac{x-b}{\bar{c}_1}) & 0 \\
0 & \exp(-s\frac{x-b}{\bar{c}_2})   &   0 & \exp(s\frac{x-b}{\bar{c}_2}) \\
0 & \frac{1}{\bar{\rho}_2\bar{c}_2}\exp(-s\frac{x-b}{\bar{c}_2})   &   0 & -\frac{1}{\bar{\rho}_2\bar{c}_2}\exp(s\frac{x-b}{\bar{c}_2})
\end{bmatrix}  
}_{\boldsymbol{\Phi}(x)}
\underbrace{
\begin{bmatrix}
\hat{g}_{1} \\ \hat{f}_{2} \\ \hat{f}_{1} \\ \hat{g}_{2}
\end{bmatrix},
}_{\hat{\mathbf{w}}\equiv
\begin{bmatrix}
\hat{\mathbf{w}}_{out} \\ \hat{\mathbf{w}}_{in}
\end{bmatrix}}
\end{align}
where $\hat{\mathbf{w}}_{out} \equiv [\hat{g}_{1}, \hat{f}_{2}]^T$ are the outgoing waves, and $\hat{\mathbf{w}}_{in} \equiv [\hat{f}_{1}, \hat{g}_{2}]^T$ are the ingoing waves. 
 The problem is closed by the acoustic boundary conditions at the inlet, $x=0$, and outlet, $x=L$, which link right and left travelling waves as 
\begin{align}\label{eq:bc1}
\hat{f}_1=R_1\hat{g}_1\exp(-s\tau_1),  \quad\quad\hat{g}_2=R_2 \hat{f}_2\exp(-s\tau_2), 
\end{align}
where $R_1$ and $R_2$ are the acoustic reflection coefficients, which may be complex, 
$\tau_1={2b}/{\bar c_1}$ and $\tau_2={2(L-b)}/{\bar c_2}$. 
By combining the jump conditions~\eqref{eq:jump1} and relations~\eqref{eq:bc1} with the characteristic decomposition \eqref{eq:wave1}, the stability of the acoustic system is governed by  a $2\times2$ nonlinear eigenvalue problem
\begin{align}\label{eq:duct_matr}
&\underbrace{
\begin{bmatrix}
1+R_1\exp(-s\tau_1)   &    -1-R_2\exp(-s\tau_2)\\
(1-R_1\exp(-s\tau_1))\left(1+\frac{\gamma-1}{\gamma\bar{p}}n\exp(-s\tau)\right)   &
\frac{\bar{c}_2}{\bar{c}_1}(1-R_2\exp(-s\tau_2))
\end{bmatrix}  
}_{\mathbf{N}_{out}(s)}
\underbrace{
\begin{bmatrix}
\hat{g}_1 \\ \hat{f}_2
\end{bmatrix}
}_{\hat{\mathbf{w}}_{out}}=\begin{bmatrix}
0 \\ 0
\end{bmatrix}, 
\end{align}
where the subscript $out$ denotes the outgoing travelling waves. 
Alternatively, by using relations \eqref{eq:bc1}, the nonlinear eigenproblem reads 
\begin{align}\label{eq:duct_matr2}
&\underbrace{
\begin{bmatrix}
 -1-R_1^{-1}\exp(s\tau_1)  &     1+R_2^{-1}\exp(s\tau_2)\\
 -\frac{\bar{c}_1}{\bar{c}_2}(1-R_1^{-1}\exp(s\tau_1))\left(1+\frac{\gamma-1}{\gamma\bar{p}}n\exp(-s\tau)\right)  &
-(1-R_2^{-1}\exp(s\tau_2)) 
\end{bmatrix}  
}_{\mathbf{N}_{in}(s)}
\underbrace{
\begin{bmatrix}
\hat{f}_1 \\ \hat{g}_2 
\end{bmatrix}
}_{\hat{\mathbf{w}}_{in}}=\begin{bmatrix}
0 \\ 0
\end{bmatrix}, 
\end{align}
where the subscript $in$ denotes the ingoing travelling waves. 
The primitive variables, and their spatial variation, can be recovered from the characteristic decomposition in Eq.~\eqref{eq:wave1}.
%
%
%=================================
%SEC:CA
%=================================
\section{Adjoint equations}
Adjoint methods enable the accurate and efficient calculation of the sensitivity of a quantity of interest to the design parameters. 
The adjoint system can be obtained by two strategies, which will be referred to as {\it continuous} and {\it discrete adjoints}~\cite[e.g.,][]{Magri2013,Aguilar2017,Magri2019_amr}. 
\subsection{Continuous adjoint approach}
A continuous adjoint approach provides the adjoint partial differential equations with respect to a sesquilinear form. Here, the following inner product  
$\left[a(x,t), b(x,t)\right]  \equiv \int_0^{T}\left(\int_0^{b-}a^*b \;dx +  \int_{b+}^La^*b \;dx\right)dt$ 
is employed as a sesquilinear form,  
where $^*$ is the complex conjugate. $a$ and $b$ represent 
generic complex functions. 
The adjoint  partial differential equations
are defined via the Lagrange-Green identity 
$\left[ \underline{\hat{q}}^+,\mathcal{N}\underline{\hat{q}}\right] = \left[ \mathcal{N}^+\underline{\hat{q}}^+,\underline{\hat{q}}\right]$. On integrating by parts,  the adjoint partial differential equations read 
\begin{subequations}
\begin{align}
\frac{\partial{u}^+}{\partial t} + \bar{c}^2\frac{\partial  {p}^+}{\partial x} & = 0,\\
\frac{\partial{p}^+}{\partial t}  + \frac{\partial  {u}^+}{\partial  x} & =0.
\end{align}
\end{subequations}
To transform the adjoint equations in the Laplace domain, the adjoint modal transformation $\underline{q}^+(x,t) = \underline{\hat{q}}^+(x)\Psi(t)$ has to be found. By eliminating the time dependency from the Lagrange-Green identity  
$\left[ \underline{\hat{q}}^+\Psi(t), \underline{\hat{q}}\exp(st) \right] = \left[ \underline{\hat{q}}^+\Psi(t)\exp(s^*t), \underline{\hat{q}} \right]$, it follows that $\Psi(t)=\exp(-s^*t)$. Therefore, the adjoint modal transformation is  
$(\cdot)^+(x,t)$ $=$ $\hat{(\cdot)}^+(x)\exp(-s^*t)$. 
The change in sign of the Laplace variable physically signifies that the adjoint equations evolve backward in time, i.e., they are anticausal. The adjoint equations in the Laplace domain read 
\begin{subequations}
\begin{align}
 -s^*\hat{u}^+ + \bar{c}^2\frac{d  \hat{p}^+}{d x} & = 0,\label{eq:adlin1}\\
-s^*\hat{p}^+ + \frac{d \hat{u}^+}{d x} & =0,\label{eq:adlin2}
\end{align}
\end{subequations}
for $x\in(0,L)-\{b\}$. 
 This set of equations is hyperbolic, hence, the solution is obtained by an adjoint characteristic decomposition 
\begin{align}
\begin{bmatrix}
\hat{p}^+_i \\ \hat{u}^+_i
\end{bmatrix}
=
\begin{bmatrix}
\exp\left(-s^*\frac{x-b}{\bar c_i}\right)   &   \exp\left(s^*\frac{x-b}{\bar c_i}\right) \\
-\bar c_i\exp\left(-s^*\frac{x-b}{\bar c_i}\right)   & \bar c_i\exp\left(s^*\frac{x-b}{\bar c_i}\right)
\end{bmatrix}  
\begin{bmatrix}
\hat{g}^+_{i,c} \\ \hat{f}^+_{i,c}
\end{bmatrix}. 
\label{eq:adwave1}
\end{align}
The subscript $c$ stands for ``continuous'' (adjoint). 
Equation~\eqref{eq:adwave1} is re-arranged as
\begin{align}\label{eq:bigPhic}
\underbrace{
\begin{bmatrix}
\hat{p}^+_1 \\ \hat{u}^+_1 \\ \hat{p}^+_2 \\ \hat{u}^+_2
\end{bmatrix}
}_{\boldsymbol{\hat{\mathbf{q}}}^+_c(x)}
= 
\underbrace{
\begin{bmatrix}
\exp(s^*\frac{x-b}{\bar{c}_1})   &   0 & \exp(-s^*\frac{x-b}{\bar{c}_1}) & 0 \\
\bar{c}_1\exp(s^*\frac{x-b}{\bar{c}_1})   &   0 & -\bar{c}_1\exp(-s^*\frac{x-b}{\bar{c}_1}) & 0 \\
0 & \exp(-s^*\frac{x-b}{\bar{c}_2})   &   0 & \exp(s^*\frac{x-b}{\bar{c}_2}) \\
0 & -\bar{c}_2\exp(-s^*\frac{x-b}{\bar{c}_2})   &   0 & \bar{c}_2\exp(s^*\frac{x-b}{\bar{c}_2})
\end{bmatrix}  
}_{\boldsymbol{\Phi}^+_c(x)}
\underbrace{
\begin{bmatrix}
\hat{f}^+_{1,c} \\ \hat{g}^+_{2,c} \\ \hat{g}^+_{1,c} \\ \hat{f}^+_{2,c}
\end{bmatrix},
}_{\hat{\mathbf{w}}^+_c\equiv
\begin{bmatrix}
\hat{\mathbf{w}}^+_{c,out} \\ \hat{\mathbf{w}}^+_{c,in}
\end{bmatrix}}
\end{align}
where $\hat{\mathbf{w}}^+_{c,out} \equiv [\hat{f}^+_{1,c}, \hat{g}^+_{2,c}]^T$ are the outgoing adjoint waves, and $\hat{\mathbf{w}}^+_{c,in} \equiv [\hat{g}^+_{1,c}, \hat{f}^+_{2,c}]^T$ are the ingoing adjoint waves. 
At the heat-source location, the adjoint momentum and energy equations in integral form  provide the adjoint jump conditions~\cite{Aguilar2017}
\begin{align} 
\llbracket \hat{u}^+ \rrbracket^{b^+}_{b^-}  = 0, \label{eq:adjump1}\quad\quad
\frac{\gamma\bar{p}}{\gamma-1}\llbracket \hat{p}^+ \rrbracket^{b^+}_{b^-} 
 = -n\hat{p}^+(b^+)\exp(-s^*\tau). 
\end{align}
 The problem is closed by the adjoint boundary conditions at the inlet, $x=0$, and outlet, $x=L$, which link right and left adjoint waves as 
\begin{align}\label{eq:adbc1}
\hat{g}^+_{1,c}=R^{+-1}_1\hat{f}^+_{1,c}\exp(-s^*\tau_1), \quad\quad \hat{f}^+_{2,c}=R^{+-1}_2 \hat{g}^+_{2,c}\exp(-s^*\tau_2). 
\end{align}
Zeroing the adjoint boundary conditions, which stem from the integration of the Lagrange-Green identity, yields the adjoint reflection coefficients at the inlet, $x=0$, and outlet, $x=L$, respectively~\cite{Aguilar2017} 
\begin{align}\label{eq:adbcs}
& R^+_1 = R^{-1*}_1, \quad\quad R^+_2 = R^{-1*}_2. 
\end{align}
The substitution of the adjoint characteristic decomposition \eqref{eq:adwave1} in the adjoint jump conditions \eqref{eq:adjump1} yields the continuous adjoint eigenproblem 
\begin{align}\label{eq:mat:ca}
\underbrace{
\begin{bmatrix}
1 + R^{+-1}_1\exp(-s^*\tau_1) &  -(1 + R^{+-1}_2\exp(-s^*\tau_2))\left(1+\frac{\gamma-1}{\gamma\bar{p}}n\exp(-s^*\tau)\right) \\
 1 - R^{+-1}_1\exp(-s^*\tau_1)& \frac{\bar{c}_2}{\bar{c}_1}(1-R^{+-1}_2\exp(-s^*\tau_2))
\end{bmatrix}
}_{\mathbf{N}^+_{c,out}(s^*)}
\underbrace{
\begin{bmatrix}
\hat{f}_{1,c}^+ \\
\hat{g}_{2,c}^+
\end{bmatrix}
}_{\hat{\mathbf{w}}^+_{c,out}}
=\begin{bmatrix}
0 \\ 0
\end{bmatrix}. 
\end{align}
Alternatively, by using relations \eqref{eq:adbc1}, the continuous adjoint  eigenproblem reads 
\begin{align}\label{eq:mat:ca1}
\underbrace{
\begin{bmatrix}
 -1 - R^{+}_1\exp(s^*\tau_1) & \left(1 + R^{+}_2\exp(s^*\tau_2)\right)\left(1+\frac{\gamma-1}{\gamma\bar{p}}n\exp(-s^*\tau)\right) \\
  -\frac{\bar{c}_1}{\bar{c}_2} \left(1-R^{+}_1\exp(s^*\tau_1)\right) & -\left(1 - R^{+}_2\exp(s^*\tau_2) \right)
\end{bmatrix}
}_{\mathbf{N}^+_{c,in}(s^*)}
\underbrace{
\begin{bmatrix}
\hat{g}_{1,c}^+ \\
 \hat{f}_{2,c}^+
\end{bmatrix}
}_{\hat{\mathbf{w}}^+_{c,in}}
=\begin{bmatrix}
0 \\ 0
\end{bmatrix}. 
\end{align}
Section 2.4 of~\cite{Aguilar2017} shows the full derivation of the continuous adjoint equations, jump conditions and boundary conditions. 
\subsection{Discrete adjoint approach}
\begin{figure}[!t]
  \begin{center}  
 \includegraphics[width=0.8\textwidth]{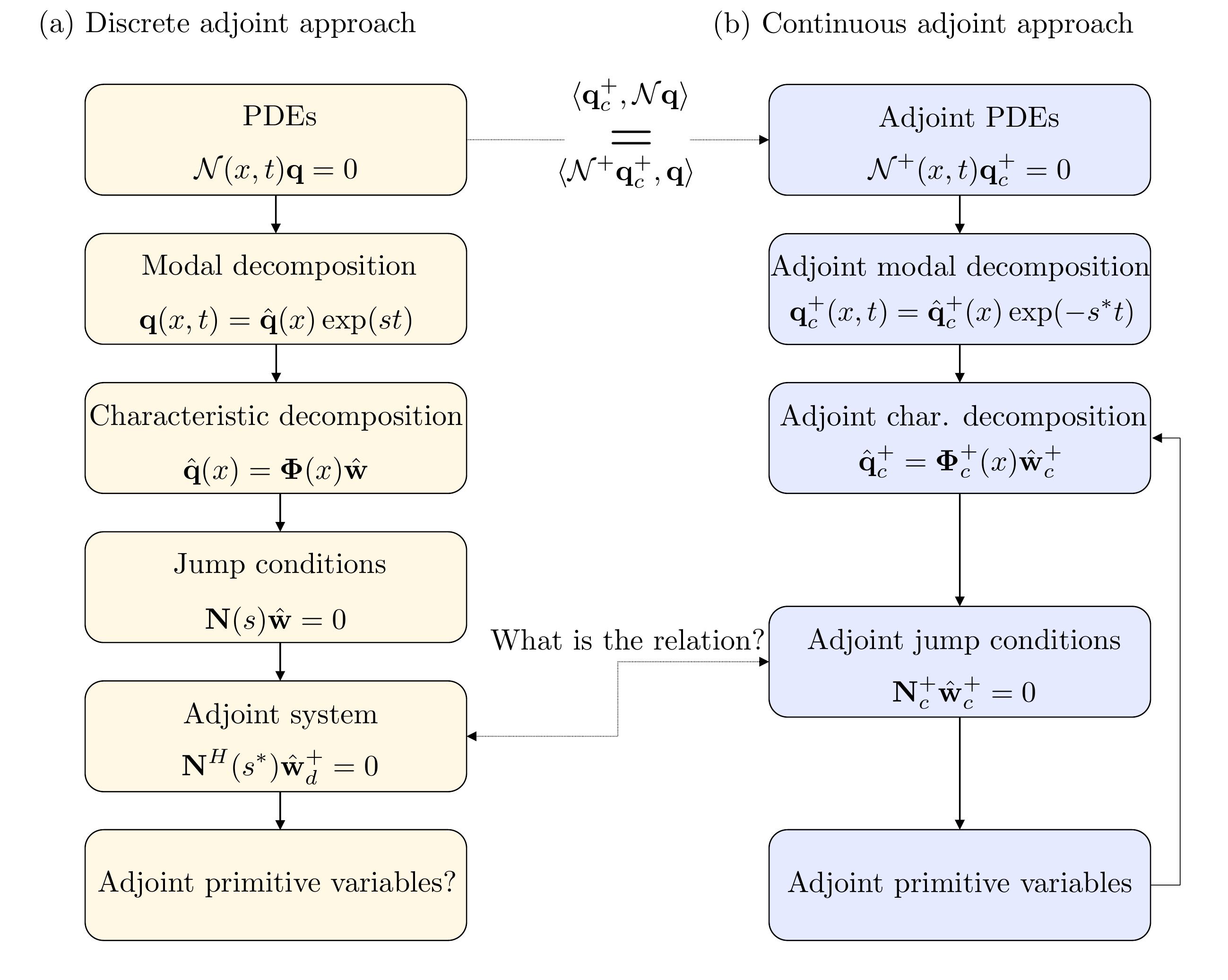}
  \end{center}
  \caption{Schematic of adjoint approaches in acoustic wave propagation. The connection between the spatial dependency of the continuous and discrete  adjoint primitive variables is not known {\it a priori}.}\label{fig:schematic_DAAD}
  \end{figure}

In the discrete adjoints, the linearized system (or algorithm) of the numerically discretized  differential equations is transposed to spawn the adjoint system. 
In one-dimensional wave propagation, the discrete adjoint approach is straightforward: The adjoint eigenproblem is obtained by taking the conjugate transpose of the direct matrix~\eqref{eq:duct_matr}  
\begin{align}\label{eq:mat_da}
\underbrace{
\begin{bmatrix}
1 + R_1^*\exp(-s^*\tau_1) &  ( 1 - R_1^*\exp(-s^*\tau_1))\left(1+\frac{\gamma-1}{\gamma\bar{p}}n\exp(-s^*\tau)\right) \\
-(1 + R_2^*\exp(-s^*\tau_2))& \frac{\bar{c}_2}{\bar{c}_1}(1-R_2^*\exp(-s^*\tau_2))
\end{bmatrix}
}_{\mathbf{N}_{out}^H(s^*)}
\underbrace{
\begin{bmatrix}
\hat{g}_{1,d}^+ \\
\hat{f}_{2,d}^+
\end{bmatrix}
}_{\hat{\mathbf{w}}^+_{d,out}}
=\begin{bmatrix}
0 \\ 0
\end{bmatrix}, 
\end{align}
where $^H$ denotes the complex conjugate transpose, and the subscript $d$ stands for ``discrete'' (adjoint).  
Alternatively, from~\eqref{eq:duct_matr2}
\begin{align}\label{eq:mat_da2}
&\underbrace{
\begin{bmatrix}
 -1-R_1^{-1*}\exp(s^*\tau_1) &  -\frac{\bar{c}_1}{\bar{c}_2}(1-R_1^{-1*}\exp(s^*\tau_1))\left(1+\frac{\gamma-1}{\gamma\bar{p}}n\exp(-s^*\tau)\right) \\
1+R_2^{-1*}\exp(s^*\tau_2) 
    &
 -(1-R_2^{-1*}\exp(s^*\tau_2)) \end{bmatrix}  
}_{\mathbf{N}^H_{in}(s^*)}
\underbrace{
\begin{bmatrix}
\hat{f}^+_{1,d} \\ \hat{g}^+_{2,d} 
\end{bmatrix}
}_{\hat{\mathbf{w}}^+_{d,in}}
=\begin{bmatrix}
0 \\ 0
\end{bmatrix}.  
\end{align}
In the discrete adjoint approach, the adjoint characteristic decomposition is not known {\it a priori}, therefore the adjoint primitive variables, and their spatial variation, cannot be calculated. 
Because the matrices from the discrete and continuous adjoint approaches are not equal to each other, i.e., $\mathbf{N}^H\not=\mathbf{N}_c^+$ (see Eqs.~\eqref{eq:mat:ca}-\eqref{eq:mat_da2}), the continuous adjoint characteristic decomposition\footnote{whose spatial dependency is provided by $\boldsymbol{\Phi}_c^+(x)$ in Eq.~\eqref{eq:bigPhic}} cannot be applied to the discrete adjoint Riemann invariants~\eqref{eq:mat_da}-\eqref{eq:mat_da2} to compute the adjoint primitive variables. %
The problem is schematically described in Fig.~\ref{fig:schematic_DAAD} and can be concisely stated as \\ 

\textbf{Problem:} \textit{What is the adjoint characteristic decomposition in the discrete adjoint approach?} \\ 

\noindent 
Section~\ref{sec:dacacon} proposes a practical method to obtain the missing spatial information in the discrete adjoint approach. 
%===============================
%SEC:CONNECTION
%===============================
% 
%
%
\section{The connection between continuous and discrete  adjoints}\label{sec:dacacon}
An eigenvalue, $\sigma$, is the solution of the transcendental equation $\det(\mathbf{N}(\sigma))=0$. 
Although not necessary, defective systems~\cite{Mensah2018_jsv} are not considered here for brevity, which implies that the ranks of $\mathbf{N}^H$ and $\mathbf{N}_c^+$ are equal to 1.
By inspection of Eqs.~\eqref{eq:mat:ca}-\eqref{eq:mat_da2}, the matrices $\mathbf{N}^H$ and $\mathbf{N}_c^+$ have the same trascendental characteristic equation, i.e., $\det(\mathbf{N}^H(s^*))=\det(\mathbf{N}^+_c(s^*))$, therefore, they have the same eigenvalue, $\sigma^*$, i.e., $\det(\mathbf{N}^H(\sigma^*))=\det(\mathbf{N}^+_c(\sigma^*))=0$.
Thus, matrices $\mathbf{N}^H(\sigma^*)$ and $\mathbf{N}^+_c(\sigma^*)$ are similar, which is the key property that enables the connection between continuous and discrete adjoint approaches.
The adjoint matrices are being evaluated at the eigenvalue $\sigma^*$ from now on and the dependency on it will be dropped for brevity. To find the similarity transformation, the adjoint matrices are eigendecomposed\footnote{Any other decomposition that provides a basis would work, e.g., the singular value decomposition.} as 
  \begin{align}\label{eq:similaritytransformations}
\mathbf{N}^H  = \mathbf{Q}\mathbf{\Lambda}\mathbf{Q}^{-1},\quad\quad
\mathbf{N}_c^+  = \mathbf{Q}_c\mathbf{\Lambda}\mathbf{Q}_c^{-1}, 
\end{align}
The similarity transformation~\eqref{eq:similaritytransformations} can be numerically calculated for problems with more degrees of freedom. However, in the $2\times 2$ matrix under investigation, an analytical expression can be derived: 
$\mathbf{\Lambda} = 
\begin{bmatrix}
\lambda_+&  0 \\
0 & \lambda_-
\end{bmatrix}$, 
where $\lambda_-=0$ because matrices \eqref{eq:similaritytransformations} have rank 1.  
Therefore, the second eigenvalue reads 
$\lambda_+ = \textrm{Tr}(\mathbf{N}^H)=\textrm{Tr}(\mathbf{N}_c^+)$. 
By definition of similarity, $\exists\; \mathbf{P}\in\mathbb{C}^{2\times 2}$ with its inverse such that 
\begin{align}\label{eq:similarityP}
& \mathbf{N}_c^+\mathbf{P}^{-1} = \mathbf{P}^{-1}\mathbf{N}^H. 
\end{align}
  Equation~\eqref{eq:similarityP} shows that the continuous and discrete  adjoint matrices are the same operator, which is represented in two different bases. 
Therefore 
 $\mathbf{Q}_c^{-1}\mathbf{N}_c^+\mathbf{Q}_c =  \mathbf{Q}^{-1}\mathbf{N}^H\mathbf{Q}
\implies \mathbf{N}_c^+\mathbf{Q}_c\mathbf{Q}^{-1} =  \mathbf{Q}_c\mathbf{Q}^{-1}\mathbf{N}^H$. 
By comparison with \eqref{eq:similarityP}, the similarity transformation can be expressed as 
\begin{align}\label{eq:f4rnuvwn3iv}
\mathbf{P}^{-1}=\mathbf{Q}_c\mathbf{Q}^{-1}.
\end{align} 
On working through the algebra and simplifying, it is possible to show that the similarity transformation reduces to a simple expression 
\begin{align}\label{eq:fhfhd}
&\mathbf{P}^{-1} = 
\begin{bmatrix}
\frac{{N}^H_{21}}{{N}^+_{c,21}} &  \frac{{N}^+_{c,11}-{N}^H_{11}}{{N}^+_{c,21}}\\
0 & 1
\end{bmatrix}, 
\end{align}
which is a similarity transformation for $2\times2$ matrices that have rank 1 when ${N}^+_{c,21}\neq0$. 
The similarity transformation~\eqref{eq:fhfhd} explicitly reads 
\begin{align}\label{eq:Pminus1}
\mathbf{P}_{out}^{-1}  &= 
\begin{bmatrix}
-\frac{1+R_2^*\exp(-\sigma^*\tau_2)}{1-R^{+-1}_1\exp(-\sigma^*\tau_1)} & 0\\
0 & 1
\end{bmatrix}\\
\mathbf{P}_{in}^{-1}  &= 
\begin{bmatrix}
\frac{1+R_2^{-1*}\exp(\sigma^*\tau_2)}{-\frac{\bar{c}_1}{\bar{c}_2}\left(1-R_1^{-1*}\exp(\sigma^*\tau_1)\right)}& 0\\
0 & 1
\end{bmatrix}. 
\end{align}
The relations between the discrete and continuous adjoint Riemann invariants read 
\begin{align}
\mathbf{P}_{out}^{-1}\hat{\mathbf{w}}_{d,out}^+  = \hat{\mathbf{w}}_{c,out}^+, \quad\quad
\mathbf{P}_{in}^{-1}\hat{\mathbf{w}}_{d,in}^+  = \hat{\mathbf{w}}_{c,in}^+. 
\end{align}

This brings us to the main result of this paper. \\

\begin{figure}[!t]
  \begin{center}  
 \includegraphics[width=0.9\textwidth]{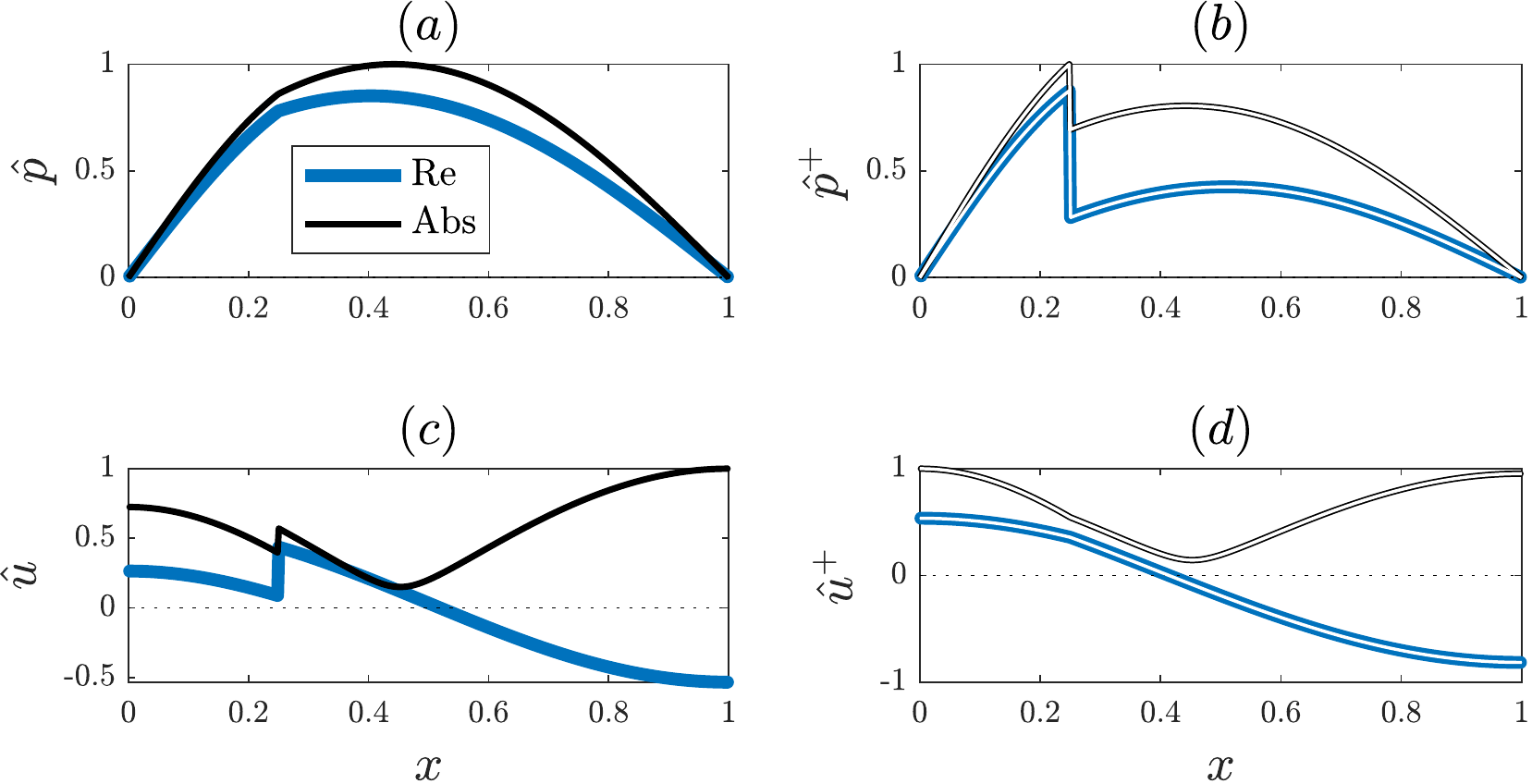}
  \end{center}
  \caption{Acoustic pressure (top) and velocity (bottom) eigenfunctions of an open-ended resonator without mean flow. Direct (left) and adjoint (right) primitive variables.  The parameters are~\cite[sec. 4.1 of ref.][]{Aguilar2017}: $L=1$ $m$, $b= 0.25$ $m$, $\gamma=1.4$, $\mathcal{R}=287.1$ $J/(kgK)$, $\bar{p}=0.1013$ $MPa$, $\bar{T}_1=300$ $K$, $\bar{T}_2=600$ $K$, $\tau=1$ $ms$, $n=1.0154\gamma\bar{p}/(\gamma-1)$, and $R_1=R_2=-1$. The eigenvalue is $\sigma=134.1$ $(1/s)$ $+$ $\textrm{i}1396.4$ $(rad/s)$. In the right panels, the white thin lines are the solutions obtained with the discrete adjoint Riemann invariants and the spatial transformation of Eq.~\eqref{eq:bigPhid}, whereas the coloured lines are the solutions calculated with a continuous adjoint approach. The pressure and acoustic eigenfunctions are normalized by $\max(|\hat{p}|)$ and $\max(|\hat{u}|)$, respectively. The adjoint pressure and acoustic eigenfunctions are normalized by $\max(|\hat{p}^+|)$ and $\max(|\hat{u}^+|)$, respectively. $\max(|\hat{p}|)/\max(|\hat{u}|)=292.1119$ and $\max(|\hat{p}^+|)/\max(|\hat{u}^+|)=0.0024$. }\label{fig:zmver}
  \end{figure}

\textbf{Result:} {\it The adjoint characteristic decomposition is given by} 
\begin{align}\label{eq:bigPhid}
\boldsymbol{\hat{\mathbf{q}}}^+_c(x)
= 
\underbrace{
\boldsymbol{\Phi}^+_c(x)
\begin{bmatrix}
\mathbf{P}^{-1}_{out} & \begin{bmatrix} 0&0\\0&0\end{bmatrix}\\
\begin{bmatrix} 0&0\\0&0\end{bmatrix} & \mathbf{P}^{-1}_{in} 
\end{bmatrix}  
}_{\boldsymbol{\Phi}^+_d(x)}
\begin{bmatrix}
\hat{\mathbf{w}}^+_{d,out} \\ \hat{\mathbf{w}}^+_{d,in} 
\end{bmatrix}. 
\end{align}
With this connection, which is numerically verified in Fig.~\ref{fig:zmver}, the first-order change of the eigenvalue, $\delta\sigma$, due to a perturbation $\mathbf{P}\delta(x-x_p)$, where $\delta(x-x_p)$ is the Dirac delta distribution centred at $x_p$,  can be calculated from the discrete adjoint approach as~\cite{Magri2016b,Magri2019_amr}
\begin{align}
\delta\sigma(x_p)=\frac{\hat{\mathbf{q}}^{+*}(x_p)\cdot\mathbf{P} \hat{\mathbf{q}}(x_p) }{\int_0^L\hat{\mathbf{q}}^{+*}(x)\cdot\frac{\partial\mathbf{N}}{\partial \sigma} \hat{\mathbf{q}}(x)\;dx}, 
\end{align}
which is useful in sensitivity analysis. 
  \subsection{Acoustics with a mean flow}
\begin{figure}[!t]
  \begin{center}  
 \includegraphics[width=0.99\textwidth]{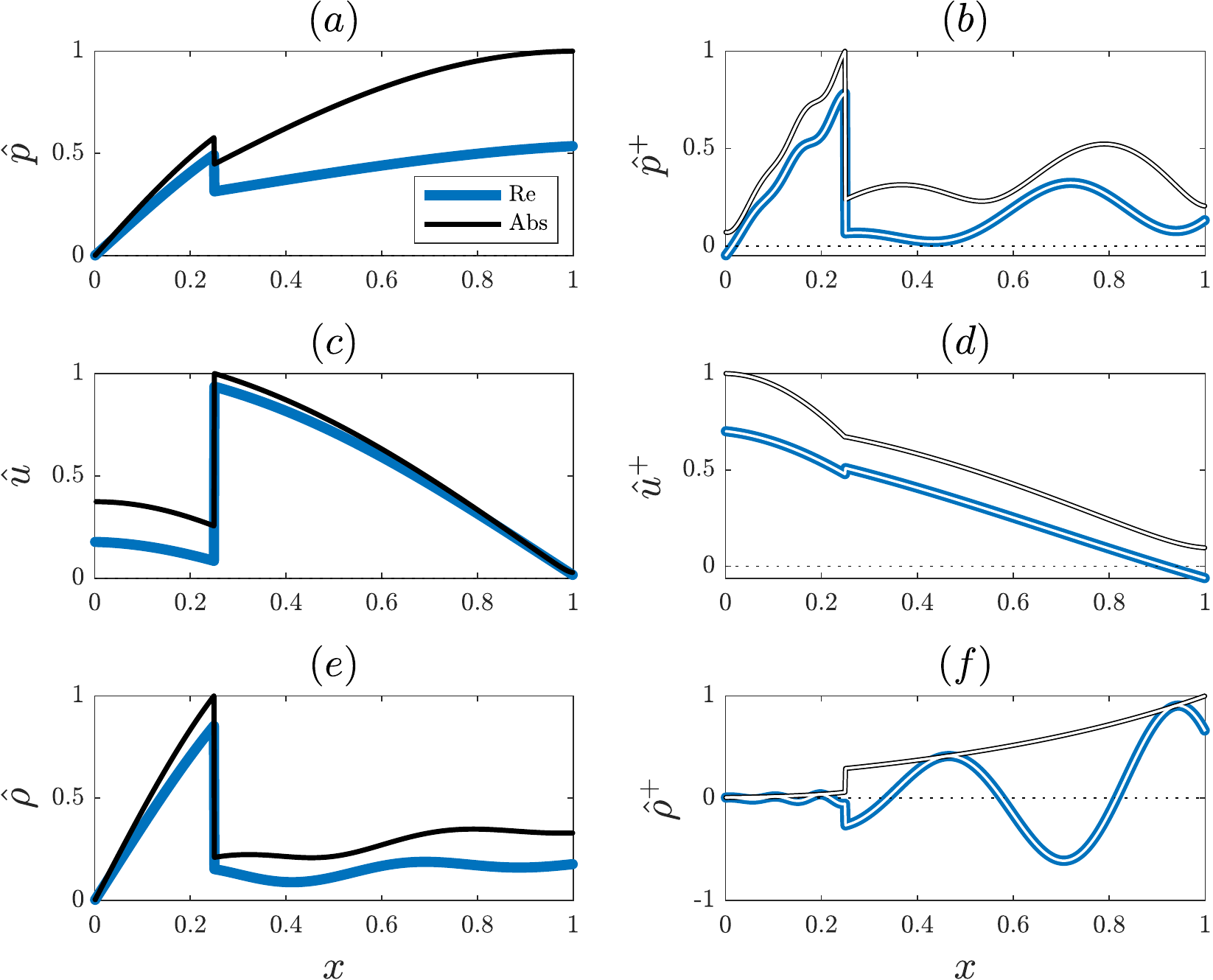}
  \end{center}
  \caption{Acoustic pressure (top), velocity (bottom), and density eigenfunctions of an open-choked resonator with a mean flow and an advected entropy inhomogeneity. Direct (left) and adjoint (right) primitive variables.  The parameters are~\cite[sec. 6.1 of ref.][]{Aguilar2017} as in Fig.~\ref{fig:zmver} with the following differences/additions: $\bar{T}_2=1500$ $K$, $\bar{M}_1=0.05$ and $\bar{M}_2=0.1135$, where $\bar{M}$ is the Mach number, $R_2=0.9556$ (choked). The eigenvalue is $\sigma=145.2$ $(1/s)$ $+$ $\textrm{i}1158.5$ $(rad/s)$. The only difference with the case of sec. 6.1 of ref.~\cite{Aguilar2017} is that the heat source is considered at rest here. The eigenfunctions are normalized similarly to Fig.~\ref{fig:zmver}, where here $\max(|\hat{p}|)/\max(|\hat{u}|)=198.6844$, $\max(|\hat{p}|)/\max(|\hat{\rho}|)=2.0885\times10^5$, $\max(|\hat{p}^+|)/\max(|\hat{u}^+|)=0.0022$ and $\max(|\hat{p}^+|)/\max(|\hat{\rho}^+|)=8.3722\times 10^{-6}$. }\label{fig:mfver}
  \end{figure}

  Adding flow inhomogeneities and mean flow effects, for which the continuous adjoint formulation becomes cumbersome, requires no conceptual modification to the similarity transformation. The full set of equations can be found in~\cite[e.g., see Sec.~5.4 in ref.][]{Aguilar2017} and~\cite{Aguilar_thesis}, which are not reported here for brevity. 
 The problem with a mean flow has three primitive variables, the pressure, $\hat{p}$, the velocity, $\hat{u}$, and the density, $\hat{\rho}$. Although an analytical derivation is still possible, the $3\times 3$ similarity matrix in Eq.~\eqref{eq:f4rnuvwn3iv} is evaluated numerically in this case. Figure~\ref{fig:mfver} shows the spatial dependence of the primitive variables and their adjoint obtained with the adjoint characteristic decomposition~\eqref{eq:bigPhid}. 
%%
%
%
%
%==============================
%SEC:Conclusions
%==============================
\section{Conclusions}
A systematic method is proposed to connect the continuous and discrete adjoint characteristic decompositions and adjoint Riemann invariants in wave propagation. The key mathematical observation is that the two adjoint operators are similar. 
To keep the mathematics simple, the method is shown in a low-order model of an acoustic resonator with a monopole source of sound.  
Because larger acoustic networks are composed of a connection of acoustic resonators,  the adjoint characteristic decomposition can be scaled up, for example, in the design of stable gas turbine combustors~\cite[e.g.,][]{Dowling2003,Lieuwen2005,Stow2009}.
With the proposed adjoint characteristic decomposition, the adjoint primitive variables can be obtained from the adjoint Riemann invariants of the discrete adjoint approach. 
Thus, the missing spatial information of the discrete adjoint approach is found. The analytical derivation is numerically verified for a case without mean flow and a case with a mean flow. 
Such spatial information enables the optimal passive control of acoustic oscillations by exploiting the discrete adjoint approach, which is versatile, straightforward to implement and provides sensitivities that are accurate to machine precision. 
This paper provides the foundation to tackle larger acoustic networks. 
The connection between continuous and discrete adjoints can be used in adjoint-based design of other problems that can be solved by the method of characteristics. 
\section*{Acknowledgements}
Financial support from the Royal Academy of Engineering Research Fellowships Scheme is gratefully acknowledged. The author thanks J. Aguilar for fruitful discussions.
%
%
%==============================
%SEC:Bibliography 
%==============================
\section*{References}

\bibliographystyle{elsarticle-num}

\end{document}